\newif\ifsubmode
\submodefalse
\submodetrue

\documentclass[12pt,preprint]{aastex}

\usepackage{natbib}

\ifsubmode
\else
\slugcomment{\today}
\fi

\def\SNR{S/N}                   
\def\FWHM{W}                   
\def\tempcompl{{\cal E}_t}                   
\newcommand{\lta}{\lesssim}
\newcommand{\gta}{\gtrsim}
\def\arcdeg{\hbox{$^\circ$}}

\shorttitle{SuperMACHO Microlensing Survey Discrimination}
\shortauthors{Rest et al.}

\begin{document}

\title{Testing LMC Microlensing Scenarios: 
The Discrimination Power of the SuperMACHO Microlensing Survey}

\author{A. Rest\altaffilmark{1,2}, C. Stubbs\altaffilmark{3}, 
A. C. Becker, G. A. Miknaitis, A. Miceli, R. Covarrubias, and S. L. Hawley}
\affil{Department of Astronomy, University of Washington, 
Box 351580, Seattle, WA 98195}

\author{R. C. Smith, N. B. Suntzeff, K. Olsen, J. L. Prieto, and R. Hiriart}
\affil{Cerro Tololo Inter-American Observatory (CTIO), La Serena, Chile}

\author{D. L. Welch}
\affil{Department of Physics and Astronomy, McMaster University,
Hamilton, Ontario, L8S 4M1, Canada}

\author{K. H. Cook, S. Nikolaev, M. Huber, and G. Prochtor\altaffilmark{4}}
\affil{Lawrence Livermore National Laboratory,
7000 East Ave., Livermore, CA 94550}

\author{A. Clocchiatti\altaffilmark{5} and D. Minniti\altaffilmark{6}}
\affil{Department of Astronomy, Pontificia Universidad Cat\'olica de Chile,
Casilla 306, Santiago 22, Chile}

\author{A. Garg, and P. Challis\altaffilmark{7}}
\affil{Physics Department, Harvard University, 17 Oxford Street, Cambridge, MA 02138}

\and

\author{S. C. Keller and B. P. Schmidt}
\affil{Research School of Astronomy and Astrophysics, Australian
National University, Weston, ACT 2611, Australia}

\altaffiltext{1}{Now at Cerro Tololo Inter-American Observatory
(CTIO), La Serena, Chile. CTIO is a division of the National Optical
Astronomy Observatory (NOAO)}

\altaffiltext{2}{Goldberg fellow}

\altaffiltext{3}{
Now at Departments of Astronomy and of Physics, 
Harvard University, Cambridge MA 
}

\altaffiltext{4}{
Now at Astronomy Dept., UC Santa Cruz, Santa Cruz, CA 95064
}

\altaffiltext{5}{
supported from FONDECYT grant 1000524
}

\altaffiltext{6}{
supported by Fondap Center for Astrophysics 15010003
}

\altaffiltext{7}{Harvard-Smithsonian Center for Astrophysics, 60 Garden St., 
Cambridge, MA 02138.}

 
\begin{abstract}

Characterizing the nature and spatial distribution of the lensing
objects that produce the previously measured microlensing optical
depth toward the Large Magellanic Cloud (LMC) remains an open problem.
We present an appraisal of the ability of the SuperMACHO Project, a
next-generation microlensing survey directed toward the LMC, to
discriminate between various proposed lensing populations.  We
consider two scenarios: lensing by a uniform foreground screen of
objects and self-lensing by LMC stars.  The optical depth for
``screen-lensing'' is essentially constant across the face of the LMC,
whereas the optical depth for self-lensing shows a strong spatial
dependence. We have carried out extensive simulations, based upon
data obtained during the first year of the project, to assess
the SuperMACHO survey's ability to discriminate between these two
scenarios.  In our simulations we predict the expected number of
observed microlensing events for various LMC models for each of our
fields by adding artificial stars to the images and estimating the
spatial and temporal efficiency of detecting microlensing events using
Monte-Carlo methods.  We find that the event rate itself shows
significant sensitivity to the choice of the LMC luminosity function, 
limiting the conclusions which can be
drawn from the absolute rate. If instead we determine the {\it
differential} event rate across the LMC, we will decrease the impact
of these systematic biases and render our conclusions more
robust. With this approach the SuperMACHO Project should be able to
distinguish between the two categories of lens populations.
This will provide important constraints on the nature of the lensing
objects and their contributions to the Galactic dark matter halo.

\end{abstract}




\section{Introduction}

An elegant way to further our understanding of dark matter halos and
to search for astrophysical dark matter candidates is to utilize the
defining feature of the dark matter:  the effect of its gravitational
field. 
\cite{Paczynski86} suggested searching for dark matter (DM) in the
form of MAssive Compact Halo Objects (MACHO's) using gravitational
microlensing.  Several groups followed this suggestion and established
microlensing searches toward the Large Magellanic Cloud (LMC) and
other nearby galaxies.  The MACHO group reported 13-17 microlensing
events toward the LMC
\citep{Alcock00a} with event timescales ranging between 34 and 230
days. They estimated the microlensing optical depth toward the LMC to
be $\tau = 1.2^{+0.4}_{-0.3}\times 10^{-7}$.  If we assume that MACHO's
are responsible for this optical depth, then a typical halo model
allows for a MACHO halo fraction of 20\% (95\% confidence interval of
8\%-50\%) with MACHO masses ranging between 0.15 and 0.9 $M_{\odot}$.
The OGLE collaboration also reported one LMC microlensing event
\citep{Udalski97}. Notably, none of the surveys toward the LMC have
detected events with timescales $1 hr \le \hat{t}
\le 10 days$. This lack of short timescale events puts a
strong upper limit on the abundance of low-mass DM objects:  objects
with masses $10^{-7} M_{\odot}<~m<~10^{-3} M_{\odot}$ make up less
than 25\% of the halo dark matter.  Further, less than 10\% of a
standard spherical halo is made of MACHO's in the $3.5 \times 10^{-7}
M_{\odot}<m<4.5 \times 10^{-5} M_{\odot}$ mass range \citep{Alcock98}.
It is interesting to note that early results from surveys toward M31 are
confirming a non-negligible MACHO content in M31's halo (\citealp{2004ApJ...612..877U};
\citealp{2005astro.ph..4188C}; \citealp{Jetzer04}).

Recent data and publications have reinvigorated discussion concerning
microlensing rates toward the LMC.  The EROS-2 project has found
evidence for variability in an event classified as microlensing by the
MACHO project \citep{2005astro.ph..1584T} while also reporting 4 new
microlensing candidates toward the LMC.  Additionally, using neural
networks to analyze a subset of MACHO lightcurves including all events
classified as microlensing by MACHO, \cite{2004astro.ph.11222E} find an
optical depth toward the LMC more consistent with that expected from
known stellar populations. \cite{Bennett05}, however, show that previously
unpublished data suppport the  microlensing interpretation questioned by
Evans and Belokurov.
Accounting for new evidence that removes
some MACHO events from the microlensing set,
\cite{2005astro.ph..2354B} recalculated the microlensing optical depth
toward the LMC using efficiencies determined for the entire data set.
His revised optical depth is $\tau = 1.0{\pm0.3}\times 10^{-7}$.  He
found the MACHO data to be consistent with a $\sim$16\% MACHO halo.
EROS-2 calculate a preliminary optical depth of about $1.5\times
10^{-8}$
\citep{Jetzer04}.  

Despite these varying constraints on the MACHO halo fraction, the
microlensing event rate (reported by \citealp{Alcock00a}) toward the
LMC significantly exceeds that expected from known visible components
of our Galaxy.  This raises the question of where the lenses reside.
Unfortunately the main observable in any given microlensing event, its
duration, depends upon a combination of lens mass, position, and
velocity relative to the source.  Any conclusion about the spatial
location of the lens population therefore depends upon the assumptions
made about its mass and velocity.  We note that in cases where the
lightcurve exhibits a departure from the point-source/point-lens event
shape (due to a binary lens, a binary source, non-inertial motion in
the lensing system, etc.), this degeneracy can be lifted.

Using the standard Galactic and LMC model, an optical depth of about
$10^{-8}$ is expected toward the LMC from known Galaxy (e.g. thick
disk, halo) and LMC components. This is significantly lower than the
optical depth of $1.2 \times 10^{-7}$ detected by the MACHO survey
\citep{Alcock00a}.  Given the difficulty associated with locating the 
lenses along the line of sight, previous microlensing surveys of the LMC 
have been unable to discriminate between a variety of possible sources
for the excess LMC event rate. These include: (1) lensing by a
population of MACHO's in the Galactic Halo, (2) lensing by a previously
undetected thick disk component of our Galaxy, (3) disk-bar or bar-bar
self-lensing of the LMC, or (4) lensing by an intervening dwarf galaxy
or tidal tail.

See \S~\ref{sec:lenspop} for a more detailed discussion of these populations.
Due to the limited number of events observed to date it is not yet
clear which scenario or combination of scenarios explains the observed
lensing signal.

The SuperMACHO Project is an ongoing five-year microlensing survey of
the LMC that is being carried out with the specific goal of answering
the question, ``Where do the lenses responsible for the measured event
rates toward the LMC reside?'' \citep{Stubbs99}.

We have designed our survey to provide a significant increase in the
number of detected events. This will allow a more robust assessment of
the spatial variation of the microlensing optical depth across the
face of the LMC, and will clarify whether the observed optical depth
can be accounted for by LMC self-lensing --- the most popular
alternative to lensing by MACHO's.
This paper presents an appraisal of SuperMACHO's ability to accomplish
this goal. This assessment is based on extensive simulations that use
observational data obtained during the first observing
season. The LMC luminosity function plays a prominent role in this
calculation, and we present an extensive analysis of this in a
future paper \citep{Rest05}.

\subsection{A First Step: Foreground lenses or LMC lenses?}

As a first step toward determining the nature of the lensing 
population, we will consider and evaluate two lensing scenarios:
\begin{itemize}
\item \textbf{\textit{Screen-lensing} :\,} Lensing caused by a
  uniform (on the angular scale of the LMC) foreground lensing
  population. Examples are lensing by the Galactic halo, thick disk, 
  or a many degree scale intervening population of lenses.
\item \textbf{\textit{LMC Self-lensing} :\,} Lensing where the
  lens population is either the same as the source population or
  spatially close to the source population. Examples are LMC
  disk-disk, disk-bar, and bar-bar lensing and lensing of the LMC
  disk by a tidal tail within the LMC.
\end{itemize}
We consider it a sensible first step to ascertain the extent to which
these may be responsible for the microlensing events seen toward the
LMC.

\begin{figure}[t]
\epsscale{0.9}
\ifsubmode
\plotone{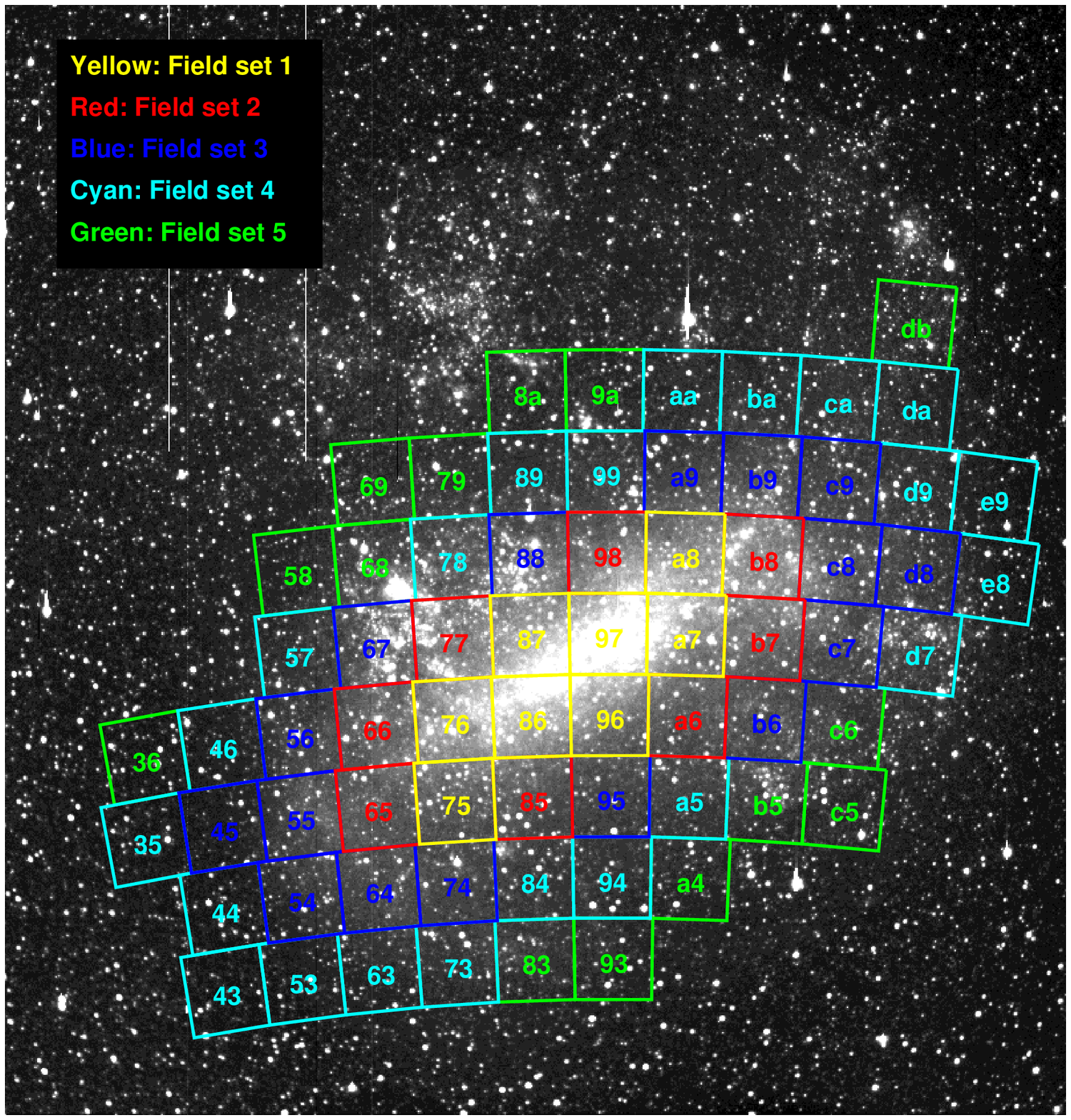}
\fi
\caption[SuperMACHO field centers]{
SuperMACHO fields superimposed onto the LMC. The fields are divided
into sets 1-5, based on their respective star density determined with
the \cite{Zhao00a} LMC bar model. Set 1 is the most crowded (yellow)
and set 5 is the least crowded set (green). (LMC image courtesy of G. Bothun.) 
\label{fig:smfields}}
\end{figure}

The lensing rate for self-lensing shows a strong spatial dependence
(e.g. the lensing rate for LMC bar-bar lensing is proportional to
$N_{bar}^2$, where $N$ is the areal density of stars), whereas the
lensing rate of screen-lensing is directly proportional to the number
of source stars observed. The goal of the SuperMACHO project is to
determine which or what mixture of these two categories causes the
observed microlensing rate. One key ingredient to achieving this is to
increase the number of detected events. 
We do this in two ways: 1) by increasing both the number of source stars monitored, 
and 2) increasing 
our event detection efficiencies by performing difference image analysis.
The corresponding improvement in event detection rate should move us out of the realm
of small number statistics and allow us to determine the spatial
distribution of the events on the sky (see
Figure~\ref{fig:smfields}). This, in turn, should allow us to 
discriminate between the two possibilities described above. 

Our approach in assessing the survey's discrimination capability is
to:
\begin {enumerate}

\item{} Use actual LMC images obtained with the survey instrumentation
to obtain star count information for each of our fields;

\item{} Add simulated microlensed flux to the frames and assess the
survey's event detection efficiency for each field as a function of
input event parameters;

\item{} Estimate, for the observed LMC optical depth, the likely event
distribution statistics across the different fields under different lensing
scenarios;

\item{} Given the anticipated event detection rates, devise statistics
that maximize the survey's ability to discriminate between screen and
LMC self-lensing; and

\item{} Assess the SuperMACHO survey's sensitivity to a specific
illustrative LMC self-lensing scenario, namely the displaced LMC bar
model proposed by \cite{Zhao00a}.
 
\end{enumerate}

These steps are laid out below. In \S~\ref{sec:observations} we
describe the SuperMACHO survey strategy and the image analysis
pipeline of the project. \S~\ref{sec:lumprof} summarizes our
parameterization of the LMC stellar luminosity function, an essential
ingredient needed to go from the observable quantity, the number of
microlensing events detected, to the physical properties of interest,
in particular the mass and spatial distributions of the lens
population. We present detailed luminosity function analysis in a
companion paper, \cite{Rest05}.  In
\S~\ref{sec:eventrateprediction}, we then use these results to
model the number of detected microlensing events for the different
candidate populations, and we predict whether it will be possible for
SuperMACHO to distinguish between screen- and self-lensing.  Before we
begin a detailed examination of SuperMACHO's discrimination
capability, we begin by examining the various lensing scenarios in
greater detail.

\section{Lens Population Candidates \label{sec:lenspop}}

The known and expected components of the Galactic/LMC system each
contribute to the event rate toward the LMC. In the following we
will discuss the different lens population candidates, each of which
is problematic in some respect.

\subsection{Milky Way Halo Lenses?}
If the lenses reside as MACHO's in the Galactic halo, their inferred
typical mass is $0.1-1\,M_{\odot}$ \citep{Alcock00a}. If we assume
that such a lens population is comprised of some known astronomical
object, the most likely candidates are white dwarfs.  There are some
indications that there might be a previously undetected population of
old white dwarfs (WDs) in the Galaxy
\citep{Ibata99,Ibata00,Mendez00,Oppenheimer01,Nelson02} favoring this
interpretation. However, \cite{Kilic04} has shown that some of these
WD candidates are background AGN.  Like any scenario populating the
Galactic halo with stellar remnants, the halo WD explanation is 
challenged by stellar formation and evolution theory: the stellar
progenitors are expected to enrich the gas and/or stars to a greater
degree than has been observed.  In addition, if other galaxies have
similar halos, then their progenitor populations should be observable
in galaxies at high redshift. There are ways out of these constraints,
e.g. by assuming non-standard initial mass functions
\citep{Chabrier96,Chabrier99}, or by anticipating lower metal yields
from old, low-metallicity main sequence progenitors, or perhaps by
allowing that the processed ejecta remain in the form of hot gas as
yet undetected \citep{Fields00}. All these attempts require fine
tuning of the models or invoke unlikely physics, rendering them
somewhat unsatisfying.  The interpretation that the detected faint WDs
are members of the Galactic halo is certainly not uncontested
(\citealp[e.g., see][]{Richer01}, \citealp{2004A&A...426...65C}).

\subsection{Thick Disk Lenses?}

An alternate interpretation is that the lenses belong kinematically
to the thick disk \citep{Reid01}. In this scenario, the inferred
number of WDs exceeds the number expected from the known thick disk or
spheroid, forcing the invocation of an undetected very thick disk as
an alternative to a halo population of lenses
\citep{Gates98}. \cite{Gates99} showed that such disks may be able to
reproduce the observed optical depth toward the LMC. More recently,
they showed that the predicted properties of such a population are
consistent with the observed properties of the WDs
\citep{Gates01}. Since the total mass of such thick disk WDs needed to
account for the observed optical depth seen toward the LMC is much
smaller than the total mass needed in the halo, this explanation
solves some of the stellar evolution and chemical enrichment
problems. Only more detailed observations can determine to which
populations the WDs might belong or if they reside in a new, unknown,
component of the Galaxy.

\subsection{LMC Self-lensing?}

A third possible interpretation of the optical depth toward the LMC is
that the lenses are not part of our Galaxy but rather of the LMC
itself, i.e. lens and source population reside within the LMC.  This
was first suggested independently by \cite{Sahu94} and \cite{Wu94} and
is denoted as LMC self-lensing.  The most common self-lensing
scenarios invoke pairs of LMC bar, disk, and halo objects as source
and/or lens populations. Several groups find
self-lensing optical depths close to $10^{-7}$ and claim that
therefore the observed optical depth can be explained with
self-lensing \citep[e.g.][]{Aubourg99,Zhao00a}.  That claim
has been disputed by several other groups
\citep[e.g.][]{Sahu94,Gould95,Alcock97a,Gyuk00} who find optical
depths closer to $10^{-8}$ and thus not sufficient to explain the
observed optical depth. The main differences between the different
estimates of the optical depth come from different choices of LMC
models and model parameters \citep{Gyuk00}.
Recent observations indicate that
there might be kinematic (carbon star sample, \citealp{Graff00};
RR Lyrae stars, \citealp{Minniti03}) and photometric subcomponents of the LMC
\citep{Weinberg01,vanderMarel01a,vanderMarel01b}. The unvirialized
subcomponents can be caused by the tidal interaction between the LMC,
the Galaxy, and/or the SMC. Some theoretical models that invoke such
unvirialized subcomponents find that the optical depth is
significantly increased and may account for half or even all of the
microlensing event rate \citep{Graff00,Zhao00a}. The predicted events
show peculiarities in their photometric, kinematic and spatial
distribution, which can be used to distinguish between LMC stars and
the other lens population candidates. For example, one of the two
near-clump MACHO events (MACHO-LMC-1) is a few tenths of a magnitude
fainter than the clump, and \cite{Zhao00b} argue that this is
suggestive of having the lensed source star in a distinct population
spatially separated and behind the LMC (and possibly more
reddened). Using Hubble Space Telescope (HST) observations,
\cite{Alcock01c} do not find any significant evidence for such
systematically redder source stars. Their results marginally favor
halo lensing.  In addition, recent observations do not show any
significant signs of kinematic outliers in the LMC \citep{Zhao02},
restricting any additional kinematically distinct population to less
than the 1\% of the LMC stars.

\subsection{Galactic Halo Substructure?}

There has been increasing evidence that the Galactic stellar halo is not
smooth. Beside the Magellanic Stream, another full-fledged tidal
stream, the Sagittarius Dwarf galaxy, which currently is passing
through the Galactic disk, has been detected
\citep{Ibata95,Yanny00,Ivezic00,Vivas01,Ibata01}. There is also
tentative evidence for other tidal streams in the Galactic Halo
\citep{Newberg02} and between the Galaxy and the LMC
\citep{Zaritsky97}.  Such an intervening dwarf galaxy or tidal tails
could also cause the high microlensing event rate
\citep{Zaritsky97,Zhao98,Weinberg01}.

\section{Observations: The Implementation of the SuperMACHO Survey}
\label{sec:observations}

The primary motivation for starting the SuperMACHO survey was the
collection of a sufficient number of microlensing events toward the
LMC so that a statistical analysis can be made of competing theories for
the location of the lenses.  Previous LMC microlensing surveys such as
MACHO and EROS have showed that microlensing can be
detected and characterized with photometric sampling every few days
and that LMC microlensing events are not shorter than two weeks.  
These surveys also
highlighted the benefits of good seeing to reduce the effects of
blending of source stars, while the simultaneous collection of data in
multiple pass-bands to assess the achromaticity of candidate events has
not proved very useful due to the effects of blending.  With these
lessons in mind, the SuperMACHO project was proposed to use a larger
aperture to detect fainter events, at a better seeing site, in a
single filter and fit into the restrictions of using a non-dedicated
telescope by observing every second night.  The image analysis was
designed to use difference image photometry which had been shown by
previous surveys to be more efficient in detecting microlensing.  The
SuperMACHO proposal \footnote{http://www.ctio.noao.edu/supermacho} was
allocated 150-half-nights, distributed over 5 years, on the Cerro
Tololo Interamerican Observatory (CTIO) Blanco 4m telescope through
the NOAO Survey Program.  The survey started in 2001 and will run
through 2005. We note that we have waived any proprietary data access
rights and that the SuperMACHO survey images are accessible through
the NOAO Science Archive on the NOAO web
site\footnote{ftp://archive.tuc.noao.edu/pub/}.

Observations are carried out every other night in dark time during the
months of October, November, and December, when the LMC is most
accessible from CTIO.  We use the $8K\times8K$ MOSAIC II CCD imager
with a FOV of 0.33 square degree.  The 8 SITe $2K\times4K$ CCDs are
read out in dual-amplifier mode (i.e.  different halves of each CCD
are read out in parallel through separate amplifiers) to increase our
observing efficiency. In order to maximize the throughput we use a
custom-made broadband filter (VR filter) from 500nm to 750nm. The
atmospheric dispersion corrector on the MOSAIC II imager allows for the
use of this broad band without a commensurate Point-Spread-Function
(PSF) degradation.

In devising an observing strategy we want to find a good balance
between maximizing the number of events detected and assuring a
uniform spatial coverage.  The work described here has guided our
decisions on how to best spend the telescope time on the sky.  We have
defined a grid of 68 fields over the face of the LMC. Previous
microlensing surveys found that the distribution of microlensing event
durations\footnote{The event duration is twice the Einstein radius crossing time} toward the
LMC has its peak at about
$\hat{t} = 80$ days, with virtually
no event lasting less than two weeks.  In order to sample the
lightcurves adequately and sufficiently, we observe all fields every
other night during dark time. This also serves to equalize, to first
order, the event detection efficiency due to sampling effects across
the fields. There remains the field-dependent detection efficiency due
to the different stellar densities and due to intentional inequality
in exposure times.
  
The distribution of the available observing time in a half-night
across the LMC is driven by two conflicting considerations: the need
to maximize the number of stars that we monitor and the need to survey
as large a region as possible in order to discriminate between the
different candidate lens populations. If the only goal were to
maximize the number of monitored stars and, consequently, the number of
detected microlensing events, we would concentrate on deep exposures
of the central region of the LMC.  Maximizing the number of
microlensing events is, however, less important than achieving maximum
discrimination between models.

We have adopted the strategy outlined by \cite{Gould99} to achieve a
distribution of exposure times that maximizes the microlensing event
rate subject to the constraints of a given spatial sampling and the
differing sensitivity between the inner and outer regions of the
LMC. The basic idea is that the distribution of exposure times for a
microlensing survey is optimized when a shift of $\delta t$ in
exposure from field A to field B gains as many stars in B as are lost
in field A. At this extremum $\frac{\delta N_A}{\delta t} =
\frac{\delta N_B}{\delta t}$.  This condition must be achieved
subject to two constraints. First, the total exposure time plus the
total time spent on readout must equal the number of workable hours in
a half night. The second constraint is that the number of sources
monitored in the inner and outer regions of the LMC must be balanced
to achieve the desired spatial coverage.

We have used the stellar density normalizations described in
\S~\ref{sec:lumprof} and a simple division into inner and outer
fields to optimize the distribution of exposure times given the
properties of the MOSAIC II imager on the CTIO 4m telescope. We have set
a minimum exposure time of 25 seconds in order to assure coverage of
the sparser fields and a maximum exposure time of 200 seconds in
order to avoid saturation effects.

The SuperMACHO Project started observations in Sept 2001. The data
analysis pipeline is currently implemented as a combination of C code,
IRAF, Perl and Python scripting tied together to provide an integrated
but modular environment \citep{Smith02}.

The first steps of the data processing, crosstalk correction and
astrometric calibration, are best done on the whole image because the
units are not completely independent. The rest of the image reduction,
as well as all of the transient analysis, breaks down naturally into
16 independent units, the amplifier-images\footnote{The MOSAIC II has 8
CCDs. Each CCD is read out by two amplifiers}, and can therefore be
efficiently handled in parallel. We employ a cluster of 18 CPUs with a
6.5 Terabyte redundant disk array.

Standard photometry of transient or variable objects becomes
inefficient in highly crowded images; therefore, we use a method
called difference image analysis (DIA) which has rapidly evolved in
the last few years. The first implementation was by
\cite{Phillips95} who introduced a method that registered images,
matched the point spread function (PSF), and matched the flux of
objects in order to detect transients.  Derivatives of DIA have been
widely applied in various projects (e.g., MACHO, \citealp{Alcock99a}; M31
microlensing, \citealp{Crotts99}; OGLE, \citealp{Wozniak00}; WeCAPP,
\citealp{Goessl02}; DLS, \citealp{Becker04}). Since the PSF varies over the
field-of-view due to optical distortions or out-of-focus images, for
example, it is essential to use a spatially varying kernel
\citep{Alard98,Alard00}.

One of the main problems with the image differencing approach is that
there are more residuals, e.g. cosmic rays and bleeds, than genuinely
variable objects in the difference image. Therefore a standard
profile-fitting software like DoPHOT \citep{Schechter93} has problems
determining the proper PSF used to perform photometry in the
difference image. When the difference image is analyzed with our
customized version of DoPHOT, we force the PSF to be the one
determined for the original, flattened image.  Applying this a priori
knowledge of the PSF helps to guard against bright false positives,
such as cosmic rays and noise peaks, which generally do not have a
stellar PSF.

All detections are added into a database.  Once the database is loaded
and objects have been identified, queries are performed on new objects
which are then classified. Objects of interest are then passed to a
graphical user interface displaying stamps from image, template, and
difference image for visual classification and interpretation.

\section{LMC luminosity functions}
\label{sec:lumprof}

In order to determine the optical depth from the observable quantities
(the number and duration of detected events), the number of potential source
stars must be known. This depends on both exposure depth and on the
luminosity function (LF) of the source star population. Knowing the LMC LF is
essential for the analysis (and prediction) of microlensing event
rates.

For this analysis we require the true LF for the stellar population
of the whole LMC.  The MACHO survey used Hubble Space Telescope (HST) images
\citep{Alcock99b,Alcock01b} to determine the LF down to
$V \sim 24$ for selected bar fields. They found that for all fields
the LF was well-fit with a power-law with identical slope for $V \lta
22.5$, leveling out for fainter magnitudes. Even the HST LF shows a
spread at magnitudes fainter than $22.5$ (see Figure 2 and 3 of
\cite{Alcock01b}). Since our survey is most sensitive to microlensing
events with source star magnitudes in the range of $22$ to $25$ (see
upper panel of Figure~\ref{fig:Nobs}), these differences are
important.  Instead of relying upon a single LF for our analysis, we
decided to explore a variety of LF's so that we can quantify the
impact of choosing an incorrect LF on our conclusions.  We used five
different LF's to represent the possible range of LF's in the LMC or
to represent possible variations in the global LF.  Two of these LF's
are the ``limiting cases'', while the three intermediate LF's are
based on either LF's from the solar neighborhood or direct fits to our
LMC photometry. As explained below, all five LF's are tied to a single
power-law fit to the bright end of the LF, and the faint-end span a
plausible range of luminosity distributions. We present our detailed
analysis of the LMC luminosity function in a companion paper
\citep{Rest05} and summarize its results here.

\begin{figure}[t]
\epsscale{0.9}
\plotone{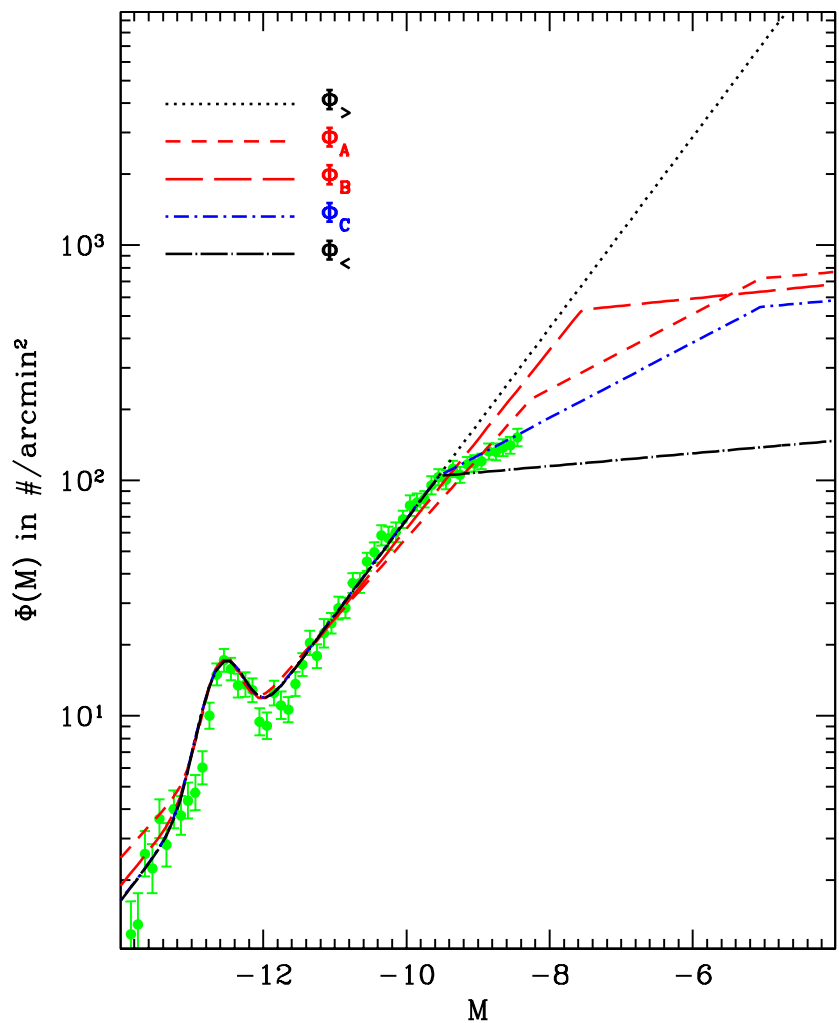}
\caption[Luminosity functions for sme9]{
This figure shows the different luminosity functions $\Phi_>$,
$\Phi_A$, $\Phi_B$, $\Phi_C$, and $\Phi_<$ for amplifier 4 of field
sme9 versus the instrumental magnitude $M$.  
The units on the y axis are stars per square arcmin and magnitude bin. 
The black dotted line is
the bright end LF function parametrized as a single power law
($\Phi_>$) fitted to the completeness corrected observed LF (green
filled circles) at the bright end.
\label{fig:LFs2sme9}}
\end{figure}

We divided each of the 68 fields that we observe into 16 subfields
based on the area covered by the MOSAIC II amplifiers (see
Figure~\ref{fig:smfields}). For each of these subfields, we determined
an independent LF. First, we fit a single power-law 
with slope $\beta$ and stellar density parameter $\Phi_0$ (stars per
square arcmin) to the bright end of the LF with a superimposed
Gaussian function representing the red clump:
\begin{eqnarray}
 \Phi^*(M) & = & \Phi_0 10^{\beta M} +
 \frac{N_{RC}}{\sigma_{RC}\sqrt{2\pi}} \exp \left[ -
 \left(\frac{M_{RC} - M}{2\sigma_{RC}} \right)^2 \right]
 \label{equ3:powerlaw}
\end{eqnarray}
where $N_{RC}$, $M_{RC}$, and $\sigma_{RC}$ are the total number, the
average magnitude, and the spread of red clump stars, respectively. 
The instrumental magnitude  $M$ is defined as
\begin{eqnarray}
M & = & -2.5 \log{f} \label{equ:M}
\end{eqnarray}
where $f$ is the flux in counts. Figure~\ref{fig:LFs2sme9} shows the
observed LF (green circles) for the magnitude range in which the
completeness is close to 100\% for amplifier 4 of field
sme9. The dotted line (denoted as $\Phi_>$) is a
completeness-corrected fit of $\Phi^*(M)$ to the data in the magnitude
range $M_{RC}-2.5 \le M \le M_{RC}+2.5$.  For fainter magnitudes ($M
\gta -9.5$) the fit clearly overestimates the number of observed
stars. This break in power-law is well known and documented for the
solar neighborhood
\citep{Kroupa93,Reid00,Chabrier01}, and it is also seen in the LMC
\citep{Holtzman97,Hunter97,Olsen99,Alcock01b}. Thus fitting a single
power-law $\Phi_>$ to the bright end of the LF's sets the stellar
density parameter for each subfield and is the upper bound to
plausible LF's.

Since the faint end deviates noticeably from the single power-law, we 
parametrize the LF's as a combination of
three power-laws:
\begin{eqnarray}
\Phi(M) & = & \left\{ \begin{array}{r@{\quad:\quad}l}
\Phi_1 10^{\beta_{1} M} & M < M_{1}\\
\Phi_1 10^{(\beta_{1} - \beta_{2}) M_{1}}  10^{\beta_{2} M} & M_{1} \le M
< M_{2}\\
\Phi_1 10^{(\beta_{1} - \beta_{2}) M_{1} + (\beta_{2} - \beta_{3}) M_{2}}
10^{\beta_{3} M} &  M \ge M_{2}
\end{array} \right. \label{equ3:LF}
\end{eqnarray}
where $M$ is defined in Equation~\ref{equ:M}.  Some of our candidate
LF's are only single or double power-laws. In these cases we
disregard the functions which do not apply. We use the fitted red
clump peak magnitude as the anchor point for the break magnitudes,
i.e. $M_{1}$ and $M_{2}$ are always with respect to $M_{RC}$.

The following is a description of how we construct the five LF's:
\begin{itemize}
\item \textbf{\textit{Upper limit LF} ($\Phi_>$):\,} The upper limit
  of the LF (black, dotted line in Figure~\ref{fig:LFs2sme9}) is set
  by assuming that the power-law $\Phi^*(M)$ (see
  Equation~\ref{equ3:powerlaw}) fitted to the bright end continues to
  the faint end without a break ($\Phi_1 \equiv \Phi_0$,
  $\beta_{1}=\beta$ for all $M$).
\item \textbf{\textit{Local neighborhood LF} ($\Phi_{A}$)
  \textit{semi-empirical mass-luminosity relation}:\,} 
  \cite{Reid02} estimates a three power-law function as one of the
  possibilities of the present-day mass function using a
  semi-empirical mass-luminosity relation.  We convert this
  present-day mass function back into a LF using the mass-luminosity
  function for lower \citep{Delfosse00,Reid02} and brighter
  \citep{Reid02} main-sequence stars, giving $\beta_{1} =0.34$,
  $\beta_{2}=0.16$, and $\beta_{3}=0.27$ with the two break magnitudes
  at $V=3.91$ and $V=7.11$. Using an unreddened absolute V magnitude
  of the red clump peak of 0.39 (Olsen, private communication), we can
  express the break magnitudes relative to the red clump as
  $M_{1}=M_{RC} + 4.3$ and $M_2 = M_{RC} + 7.5$. For each subfield we
  determine the stellar density parameter $\Phi_{A0}$ and the red
  clump peak magnitude $M_{RC}$ by fitting the single power-law
  $\Phi^*(M)$ (see Equation~\ref{equ3:powerlaw}) to the bright end of
  the LF with the fixed slope of $\beta = \beta_{1} =0.34$. This
  LF is shown as red, short-dashed line line in Figure~\ref{fig:LFs2sme9}.
\item \textbf{\textit{Local neighborhood LF} ($\Phi_{B}$)
  \textit{empirical mass-luminosity relation}:\,} 
  Another proposed form of the present-day mass function by
  \cite{Reid02} is a two power-law based on an empirical
  mass-luminosity function. In the same way as for $\Phi_{A}$, we
  estimate $\beta_{1} = 0.38$, $\beta_{2}=0.0314$, and $M_{1}=M_{RC} +
  5.0$. The stellar density parameter $\Phi_{B0}$ is determined by
  fitting the single power-law $\Phi^*(M)$ to the bright end of the LF
  with the fixed slope of $\beta = \beta_{1} =0.38$.
  This LF is shown as red, long-dashed line in
  Figure~\ref{fig:LFs2sme9}.
\item \textbf{\textit{Empirical LF} ($\Phi_C$):\,} 
   As our empirical LF $\Phi_C$, we fit the single power-law
  $\Phi^*(M)$ to the bright end of the LF, yielding $\Phi_1 \equiv
  \Phi_0$ and $\beta_{1}=\beta$. We fit a second power-law to the
  faint end of one of our sparse subfields, where the break in
  power-law is virtually unaffected by completeness (see
  Figure~\ref{fig:LFs2sme9}), yielding $M_{1} = M_{RC} + 3.0$ and
  $\beta_2 = 0.16$. Note that the break 3 magnitudes fainter than the
  red clump peak magnitude is in very good agreement with the break
  seen in HST images of the LMC \citep[see][]{Alcock01b}.  Since we
  don't have data going deep enough to see the second break, we assume
  the same break magnitude and slope for the third power-law $M_{2} =
  M_{RC} + 7.5$ and $\beta_3 = 0.027$, respectively, as determined for
  $\Phi_A$.
  This LF is shown as blue, dot-dashed line  in
  Figure~\ref{fig:LFs2sme9}.
\item \textbf{\textit{Lower limit LF} ($\Phi_<$):\,} 
  Similar to $\Phi_>$ and $\Phi_C$, we determine $\Phi_1$ and
  $\beta_{1}$ by fitting the single power-law $\Phi^*(M)$ to the
  bright end of the LF. As the first break in power-law, we use the
  same break we found for $\Phi_C$: $M_{1} = M_{RC} + 3.0$. As the
  slope we choose the smallest slope of any of the power-laws,
  $\beta_2=0.027$. This choice underestimates the LF at the faint
  end.  This LF is shown as black, dot-dashed line in
  Figure~\ref{fig:LFs2sme9}.
\end{itemize}
An overview of the different LF's is given in Table~\ref{tab:LFoverview}.
In a future paper \citep{Rest05} we present a more detailed description
on how we derive the LF's.

We will use these different trial luminosity functions in the following
section where we predict event rates in different microlensing scenarios. 
We show that a differential rate analysis of the data can discriminate 
between models independent of the actual underlying luminosity function.
 
\section{Event Rate Prediction}
\label{sec:eventrateprediction}

The SuperMACHO Project's initial goal is to distinguish between two
broad categories of lensing: screen-lensing and self-lensing. We use
data from the first year to predict the number of detected
microlensing events from the different candidate populations in order
to test if the SuperMACHO Project is able to distinguish between
them. We add artificial stars to the images and determine the
completeness of the detections. With that we estimate the spatial and
temporal efficiency of detecting microlensing events in our data sets.
In combination with the stellar luminosity functions described above,
we predict and compare the number of microlensing events for various
screen-lensing and self-lensing scenarios.

The number of observed events depends on both the source and lens
populations.  
In general, estimating the number of detectable microlensing events
$N_{ml}$ is complicated as it requires detailed knowledge about number
density, velocity distribution, and other properties of the
population. A good approximation is given by:
\begin{eqnarray}
  N_{ml} & = & N_{obs} \, \tau \, \tempcompl \label{equ:Nml}
\end{eqnarray}
where $N_{obs}$ is the number of monitored stars for which
microlensing can be detected, $\tau$ is the microlensing optical
depth, and $\tempcompl$ is the sampling efficiency.  This
approximation separates the photometric and temporal completeness
which makes the calculation much simpler. The number of monitored
stars contains the photometric completeness and is
\begin{eqnarray}
 N_{obs} & = & \int \Phi(M) {\cal E}(M)\, dM \label{equ:Nobs}
\end{eqnarray}
where ${\cal E}(M)$ is the efficiency of detecting microlensing for
a star with magnitude $M$, and $\Phi(M)$ is the luminosity function. The
temporal completeness is contained in $\tempcompl$ as
\begin{eqnarray}
\tempcompl & = & \left( \int  \left( \frac{\hat{t}}{T}\right) \, \frac{D(\hat{t})}{P_T(\hat{t})}  \, d\hat{t} \right)^{-1}\label{equ:tempcomp}
\end{eqnarray}
where $T$ is the effective survey duration, $\hat{t}$ is the duration of
a microlensing event, $P_T(\hat{t})$ is the probability that the
event is detected within $T$ given the temporal cadence of the survey,
and $D(\hat{t})$ is a normalized distribution of $\hat{t}$.

We calculate $N_{obs}$ and $\tempcompl$ independently for each field
and amplifier, as described below.

\subsection{Number of Observed Stars}
\label{sec:calcNobs}

The best way to estimate the number of observed stars is to perform
Monte Carlo simulations with all the images used. The shortcoming of
this method is that it is very CPU-intensive and terabytes of images
have to be simultaneously available on disk. Our goal is to predict a
lower limit on the number of events; therefore, we choose a slightly
different approach and perform Monte Carlo simulations on only a subset
of images subject to certain assumptions.

The question we have chosen to answer is: What is the probability
${\cal E}(M)$ that the flux of a star with a given magnitude $M$ is
magnified during a microlensing event by a detectable amount?  Let us
assume that a microlensing event is detectable if the difference flux
at peak has a signal-to-noise ratio $\SNR \ge 5$. Whether such an
event is then indeed detected depends on the temporal cadence as well
as the seeing and transparency of the observing nights. We fold this
into the temporal completeness analysis in \S~\ref{sec:tempcompl}.

During a microlensing event, the source star flux $f_0$ gets amplified
to $f = f_0 * A$ with an amplification that can be expressed as $A(u)
= (u^2+2)/(u \sqrt{u^2+4})$, where $u$ is the angular separation
between source and lens in units of the Einstein angle.  With
difference imaging, the quantity we measure is the not the total
amplified flux, but the difference in flux, $\Delta f = f - f_0$,
between the amplified and unamplified source star.  By measuring only
the difference flux, we avoid confusion due to crowding and blended
sources.  Using the difference images, then, we construct lightcurves
of the difference flux, $\Delta f$.  

Using Equation~\ref{equ:M}, we
can then express this flux difference as an instrumental magnitude
$M_{diff}$ where,
\begin{eqnarray}
M_{diff} (M,u) & = & - 2.5 \log (\Delta f) \\
               & = &  M - 2.5 \log (A(u) - 1 )
\end{eqnarray}
We note that this is the instrumental magnitude of the flux difference,
which is {\it not} the difference in magnitude between the amplified
and unamplified source star.

In \S~\ref{sec:results} we show that nearly all of the expected
microlensing events have source stars in the magnitude range of $22.3
< M_{VR} < 25.3$. Thus we can assume that the unamplified source star
flux does not contribute significantly to the noise in the photometry,
i.e. the photometry is background noise dominated.  Within this limit,
then, instead of adding the unamplified flux $f_0$ to the template
image, and the amplified flux $f$ to the image, we can just add the
flux difference $\Delta f$ to the detection image to test if we can
detect such a flux difference. For this simulation we add flux into
the images only and process these images through our difference image
pipeline in the standard way.  Because there is no source flux in the
template, we measure the added difference flux, $\Delta f$, in the
resultant difference image.  We then derive what fraction of the
artifical stars we recover with a $\SNR \ge 5$ to obtain the empirical
completeness function $C(M_{diff})$. This allows us to estimate the
probability that a microlensed star with intrinsic magnitude $M$ has a
change in flux whose $\SNR \ge 5$ at maximum amplification:
\begin{eqnarray}
{\cal E}(M) & = & \int C(M_{diff}(M,u_0))\,du_0 \label{equ:eps}
\end{eqnarray}
where $u_0$ is the angular separation at maximum brightness.  With
Equation~\ref{equ:Nobs}, the number of observed stars $N_{obs}$ can
then be calculated using the luminosity function $\Phi(M)$ for a given
field and amplifier (see \S~\ref{sec:lumprof}).

\subsection{Sampling efficiency}
\label{sec:tempcompl}

The temporal cadence and observing conditions such as seeing and
transparency have a significant impact on event detection efficiency. This is
folded into $P(\hat{t})$ and thus implicitly into $\tempcompl$ (see
Equation~\ref{equ:tempcomp}). 

\begin{figure}[t]
\epsscale{0.9}
\plotone{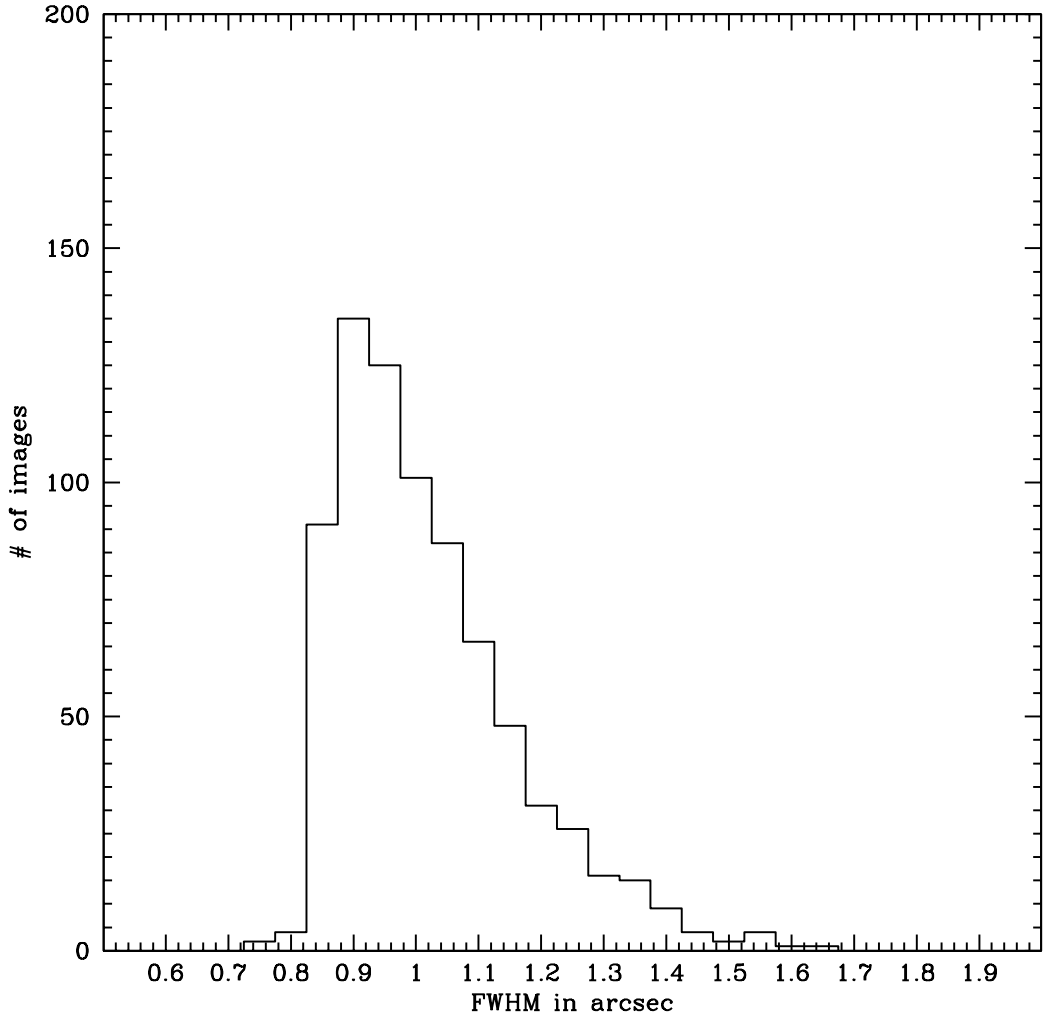}
\caption[FWHM histogram]{
Seeing (FWHM) histogram for the 2001/2002 run for amplifier 4. The average is
$1.02"$ with a standard deviation of $0.14$.
\label{fig:fwhm}}
\end{figure}

In order to estimate $P(\hat{t})$ we perform Monte-Carlo
simulations. During each year microlensing events are drawn
at random to have peak amplification sometime during an interval of
$T$ = $300$ days. This interval is dictated by the extent of the period of
actual observing (100 days) and the need to ensure that real events
which reach peak amplification before or after that interval (but
are observable during the survey months) are represented in the simulation. 
The 100-day padding at both ends of the observation period is driven
by the characteristic $\hat{t}$ found by the MACHO Project. 
The SuperMACHO observation periods are usually allocated in three runs 
of about ten nights of bi-nightly observing separated by two weeks of bright time. 
Factors such as weather, instrument failures, and computer
down time are taken into account by randomly eliminating one out of
four nights.  For a given $\hat{t}$, we realize 1000 lightcurves for
each microlensing impact parameter 
$0<u_0<0.5$ in increments of 0.0125.
By choosing $u_0=0.5$ as our upper limit we ensure that we have at
least a magnification by a factor of 2.18. This will allow us to
eliminate astrophysical sources of low-amplitude variability.  Because
the optical depth is typically calculated for $u_0\le1$, this
simulation will only recover 50\% of the number of events 
usually associated with a given optical depth The time of
maximum is randomly chosen within $T$. As a lower limit, we assume
that all microlensing events observed have a difference flux with
$\SNR\left( \FWHM_0,t_0 \right)=5$, where $\FWHM_0$ is the $FWHM$ of
the seeing at time $t_0$ of maximum amplification. Then the $\SNR$ of
a detection on another night
at time $t$ 
can be estimated as
\begin{eqnarray}
\SNR \left( \FWHM,t \right) & = & \left(\frac{\FWHM_0}{\FWHM}\right)^2 \frac{A(t)-1}{A_{max}-1} \SNR \left( \FWHM_0,t_0 \right)
\end{eqnarray}
We randomly draw $\FWHM_0$ and $\FWHM$ from the distribution of seeing
in the first year of the survey (see Figure~\ref{fig:fwhm}). We want to
emphasize that this is a conservative lower limit, since the stars
we count in $N_{obs}$ have a $\SNR \ge 5$ in the difference flux at
maximum amplification; whereas, our Monte-Carlo simulations assume
$\SNR = 5$ at peak.

For each lightcurve we estimate the $\SNR$ for every night
we have taken data based on the $FWHM$ and amplification. The decision
whether we do or do not detect such a lightcurve is based on the
following additional criteria:
\begin{itemize}
\item At least 2 detections on the rising arm of
the lightcurve with a $\SNR>2.0$
\item At least 4 detections have a $\SNR>2.0$
\end{itemize}
This ensures that the events are ``contained'' within the survey
coverage time and that there are enough significant detections to
``trigger'' an alert for follow-up observations.  
These cuts define the upper limit on the number of microlensing events
we recover.  We note the above set of criteria is insufficient to
discriminate between microlensing and the population of background
events in the actual survey.  They are, however, comparable to the
trigger criteria used in the MACHO alert system.  Here we only
consider the event sample in the case where there is no
need for further discrimination. Discriminating microlensing is a
separate problem and beyond the scope of this paper.

In the final analysis, the initial cuts described above will define a
population of candidate events.  We will apply additional cuts to
eliminate contaminants such as supernovae, AGN's, and intrinsically
variable stars.  We will use photometric and spectroscopic follow--up
observations to help define this advanced set of cuts. We note that
these cuts will lower our detection efficiencies and require a
reanalysis when they are determined.

\begin{figure}[t]
\epsscale{0.9}
\plotone{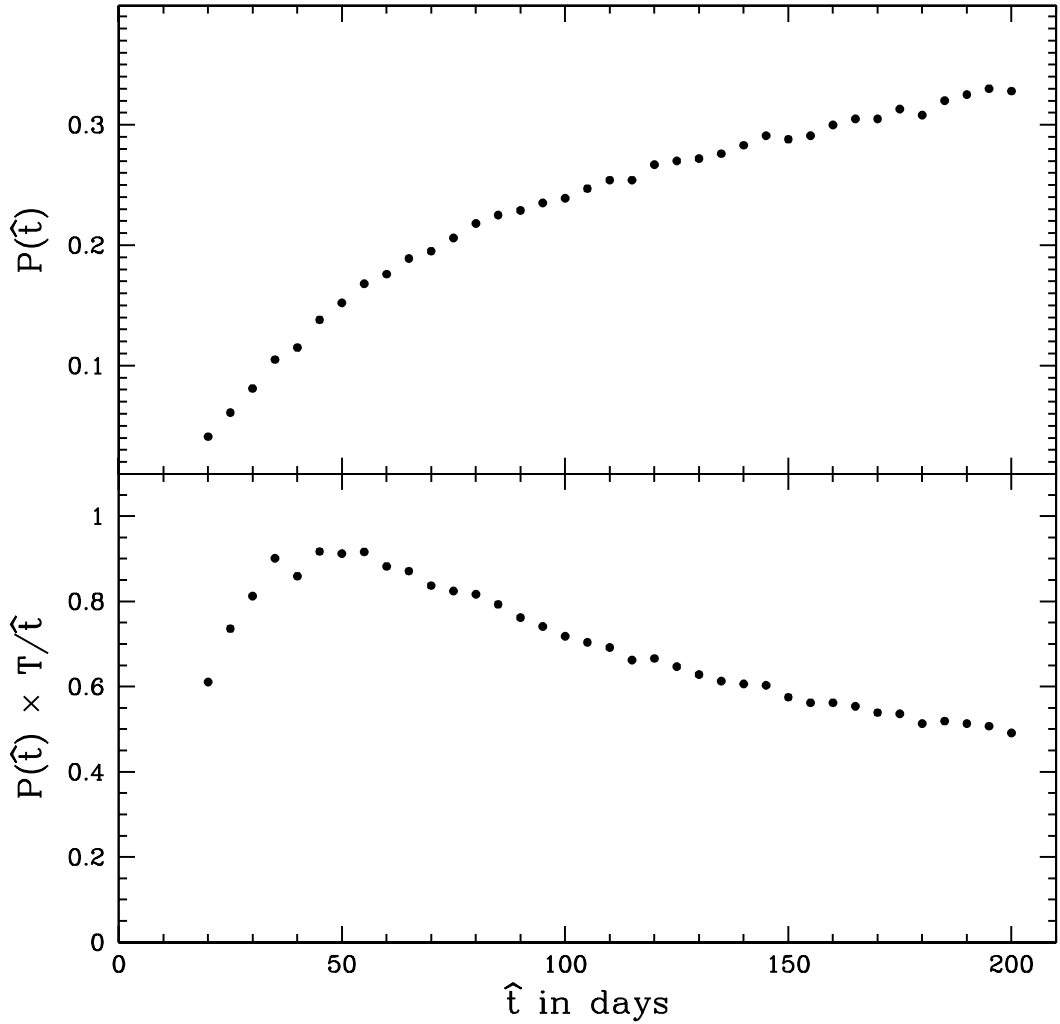}
\caption[Probability of detecting a microlensing event with event
duration of $\hat{t}$]{ Probability of detecting a microlensing event
with event duration of $\hat{t}$.  The upper panel shows the
probability $P\left(\hat{t}\right)$ of detecting a microlensing event
with event duration of $\hat{t}$. The lower panel shows the more
intuitive measure $P\left(\hat{t}\right) \times T/\hat{t}$, which is
the number of microlensing events detected per $1/\tau$ stars during
an interval $T$, assuming that all microlensing events have an event
duration of $\hat{t}$. The decrease of detection probability for
decreasing event duration is countered by that fact that for shorter
event durations more microlensing events happen per $1/\tau$ stars in
the given observing time. This causes the peak at $\hat{t} \sim 50$
days.
\label{fig:tml}}
\end{figure}

The upper panel of Figure~\ref{fig:tml} shows the probability
$P\left(\hat{t}\right)$ of detecting a microlensing event with event
duration of $\hat{t}$.  This event detection probability increases with event
duration. Note that the probability is well below $1.0$. This is
because the interval used to define the input microlensing population was
$T=300$ days, much longer than the actual annual observing duration
($\sim 100$ days), and the probability of detecting an event with a
peak time $t_0$ well outside the time the observations are taken is rather
small. This effectively cancels out later on since we multiply by $T$
when calculating $\tempcompl$ (see Equation~\ref{equ:tempcomp}).

A more intuitive measure is $P\left(\hat{t}\right) \times T/\hat{t}$,
which is the number of microlensing events detected per $1/\tau$ stars
during $T$ assuming that all microlensing events have an event
duration of $\hat{t}$ (see lower panel of Figure~\ref{fig:tml}). The
decrease of detection probability for decreasing event duration is
countered by that fact that for smaller event durations more
microlensing events happen per $1/\tau$ stars in the given observing
time. This causes the peak at $\hat{t} \sim 50$ days.

Existing microlensing event statistics suggest that the event duration
has a peak at about 80 days. Therefore we choose as $D(\hat{t})$ a
Gaussian distribution with the peak at 80 days and with a spread of 20
days as inferred from Figure 9 in
\cite{Alcock00a}.
Using Equation~\ref{equ:tempcomp}, we can now evaluate
\begin{eqnarray}
\tempcompl & \sim & 0.8 \label{equ:tempcompval}
\end{eqnarray}
Thus we will observe about 0.8 microlensing events per $1/\tau$
observed stars in one survey year (spanning 100 days). 

\subsection{Bar-Disk Self-lensing Models for the LMC}
\label{sec:selflensingmodels}

The optical depth from self-lensing remains a matter of
controversy. Some studies find rather small values in the range of
$(1.0-8.0) \times 10^{-8}$ (e.g. \citealp{Alcock97a}, \citealp{Gyuk00}, and
\citealp{Jetzer02}); whereas, other studies suggest optical depths up to
$1.5 \times 10^{-7}$ (e.g. \citealp{Zhao00a}). This controversy arises
from the still rather imprecise knowledge of the structure
of the LMC and consequent differences in the adopted models.

The \cite{Zhao00a} models derived in their paper are concrete and
testable self-lensing models which we will use in the following
sections to predict the SuperMACHO self-lensing event rate. As pointed
out above, these models, denoted as model set {\it A}, see \S~\ref{sec:zhaomodels}) 
predict a rather large optical depth and are thus more
favorable to a self-lensing interpretation of LMC microlensing.  We
then improve upon their models (see \S~\ref{sec:nikolaevmodels})
for an alternative, more realistic model set, denoted as model set
{\it B}. This allows us to make a direct spatial comparison between
self-lensing and screen-lensing event rates.

\subsubsection{\cite{Zhao00a} Model Set {\it A}}
\label{sec:zhaomodels}

\cite{Zhao00a} derive in their paper concrete and testable
self-lensing models. They define a coordinate system with $X$, $Y$,
and $Z$ being decreasing right ascension, increasing declination, and
line-of-sight direction centered at the optical center of the LMC bar,
where $X$, $Y$, and $Z$ are in units\footnote{For the LMC, 1 kpc is
roughly 1 degree} of kpc. They approximate the average separation
between source and lens as
\begin{eqnarray}
\Delta(X,Y)  & = & \frac{I_b^2\Delta_b + I_d^2\Delta_d + I_b I_d \,\mbox{max}(\Delta_b+\Delta_d,\Delta_{bd})}
{(I_b+I_d)^2}\label{equ:deltaXYZhao}
\end{eqnarray}
where $I_d$ and $I_b$ are the star count density of the disk and bar,
respectively.  The line-of-sight depth of the bar and disk are denoted
as $\Delta_b$ and $\Delta_d$, respectively, and $\Delta_{bd}$ is the
line-of-sight separation between the mid-planes of the bar and disk. In
the limit in which the source and the lens are at roughly the same
distance, the optical depth can then be expressed as
\begin{eqnarray}
\tau(X,Y) &\sim& 10^{-7} \frac{\Sigma(X,Y)}{160\; M_{\odot}\;\mbox{pc}^{-2}} \frac{\Delta(X,Y)}{1\;\mbox{kpc}}
\label{equ:tauself}
\end{eqnarray}
The value of $\tau$ depends mainly on two parameters: the displacement
$Z_0$ between the disk and the bar and the mass of the bar defined by
the mass fraction $f_b$, which are implicit in the average separation
and surface brightness.  

In \S~\ref{sec:results}, we will vary these
parameters to estimate the spatially varying optical depth for
different model realizations, and we denote this set of models as
model set {\it A}.

\subsubsection{Modified \cite{Zhao00a} Model Set {\it B}}
\label{sec:nikolaevmodels}

The optical depth range for self-lensing found by \cite{Zhao00a} is
significantly larger than the ones found by other studies
(e.g. \citealp{Alcock97a}, \citealp{Gyuk00}, and \citealp{Jetzer02}). As an
alternative to model set {\it A}, we modify the original
\cite{Zhao00a} models by improving the effective separation
$\Delta'(X,Y)$ (Nikolaev, private communication) which significantly
decreases the optical depth:
\begin{eqnarray}
\Delta'(X,Y)  & = & \frac{I_b^2\Delta_b/6 + I_d^2\Delta_d/6 + I_b I_d \Psi(\Delta_b+\Delta_d,\Delta_{bd})}
{(I_b+I_d)^2} \label{equ:deltaXYNik}
\end{eqnarray}
where the function $\Psi$ is
\begin{eqnarray}
\Psi(\Delta_b+\Delta_d,\Delta_{bd})& = & 
\left\{ \begin{array}{r@{\quad:\quad}l}
\Delta_{bd}&\Delta_{bd} > (\Delta_b + \Delta_d)/2\\
\Delta_{bd} + ((\Delta_b + \Delta_d)/2 - \Delta_{bd})^3 / (3\Delta_{b}\Delta_{d})&
\Delta_{bd} \le (\Delta_b + \Delta_d)/2\\
\end{array} \right. \label{equ:deltabdNik}
\end{eqnarray}
We denote this model set as model set {\it B}.  Note
that the first two terms of $\Delta'(X,Y)$, the contributions of
disk-disk and bar-bar self-lensing, are smaller by a factor of six
compared to the equivalent expression in the \cite{Zhao00a}
calculations (see Equation~\ref{equ:deltaXYZhao}). This takes
geometrical considerations into account since source and lens
population are the same. The third term is a better approximation for
the disk-bar separation: The two cases in
Equation~\ref{equ:deltabdNik} represent situations 1) where disk and
bar are well separated along the line of sight and 2) where they
overlap each other. The two limiting cases produce the same result
when $\Delta_{bd} = (\Delta_b + \Delta_d)/2$.

It is also important to note that the Zhao models have unrealistically
high projected central surface densities, reaching $640
\,\mbox{M}_{\odot}\,\mbox{pc}^{-2}$. A model with parameters as in
\cite{Gyuk00} leads to central densities on the order of $300
\,\mbox{M}_{\odot}\,\mbox{pc}^{-2}$. The discrepancy is due to the
very heavy bar ($f_b=0.5$) and the quartic bar model (which is more
centrally concentrated than a Gaussian bar) used by \cite{Zhao00a}.
Less centralized projected surface densities result in a
decrease of the central self-lensing optical depth.  
As before with model set {\it A}, we vary the displacement $Z_0$ and
the mass fraction $f_b$ in \S~\ref{sec:results} and denote this set of
models as model set {\it B}.

\section{Results}
\label{sec:results}

In this section, we combine the results from the previous sections in
order to obtain a quantitative prediction of the anticipated number of 
microlensing events. The main observable difference between self-lensing and
screen-lensing is their distinctive spatial
dependence of event rate; therefore, we calculate separately for each field and
amplifier the expected number of observed microlensing events using
Equation~\ref{equ:Nml}.

\begin{figure}[t]
\epsscale{0.9}
\plotone{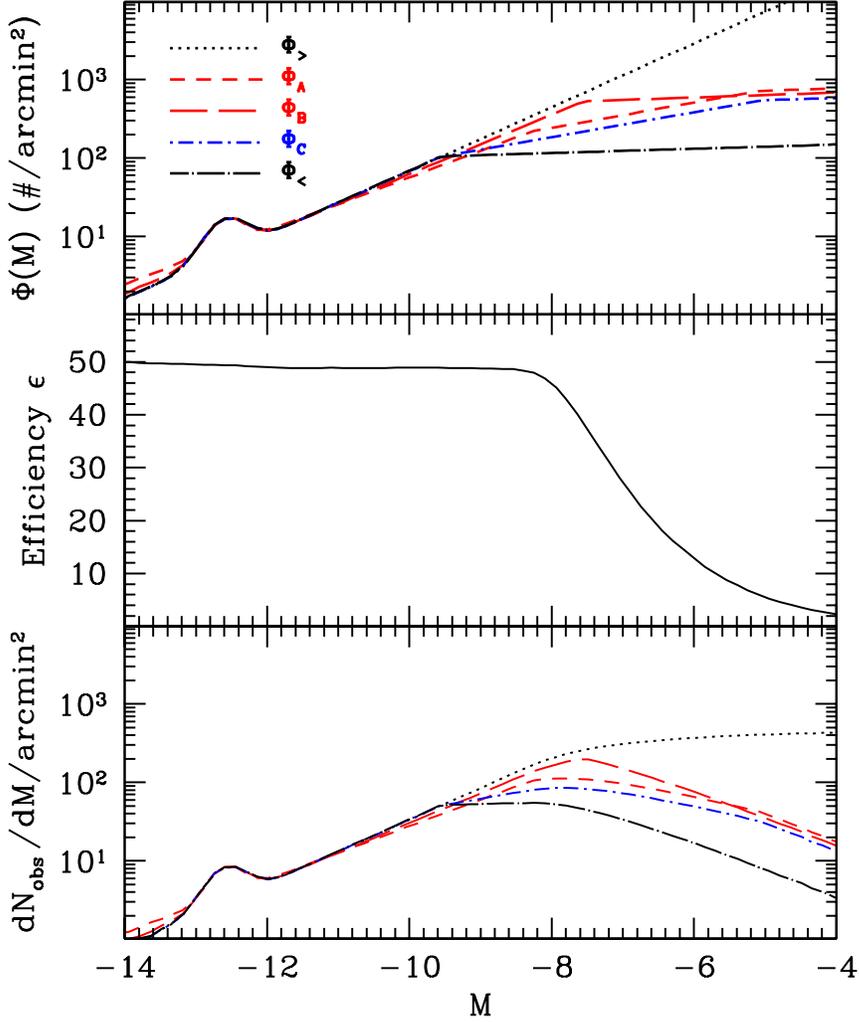}
\caption[Number of observed stars for field sme9]{
Determination of the number of observed stars in field sme9.  The
upper panel shows the LF candidates $\Phi_>$, $\Phi_A$, $\Phi_B$,
$\Phi_C$, and $\Phi_<$, constructed as described in the text. The
x-axis is the instrumental magnitude. The middle panel shows the event
detection efficiency as a function of source star magnitude. The lower
panel shows the number of monitored stars vs. magnitude, which
is the product of the two upper panels.
\label{fig:Nobs.sme9}}
\end{figure}

\begin{figure}[t]
\epsscale{0.9}
\plotone{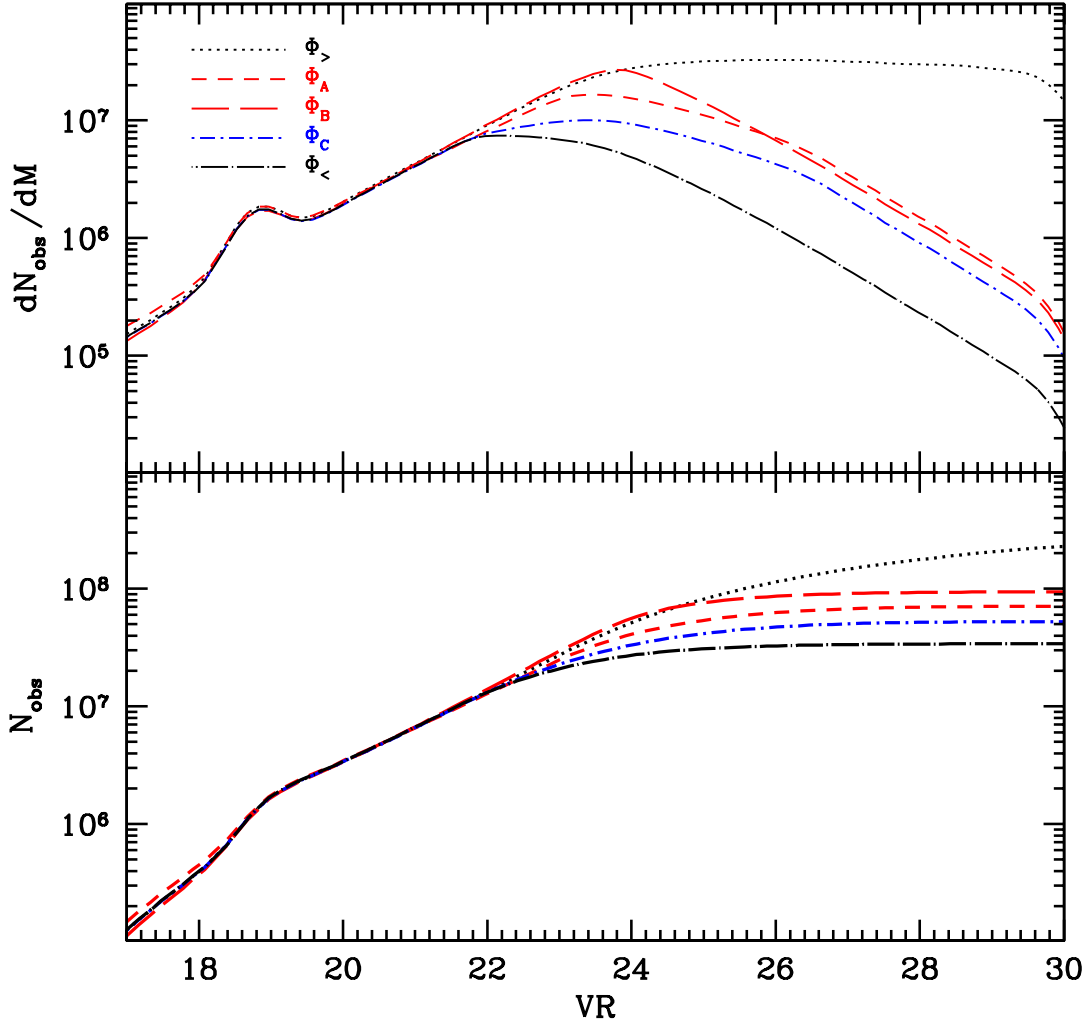}
\caption[Differential and cumulative $N_{obs}$ for different LF's]{
Differential and cumulative $N_{obs}$ for different LF's. The x-axis is
the apparent {\it VR} magnitude, synthesized from standard $V$ and $R$
magnitudes through the equation $VR=A_0 -2.5\times\log_{10}(A_1
10^{-0.4V} + (1-A_1)10^{-0.4R})$, where $A_0$ and $A_1$ are typically
-1.1 and 0.37 for the MOSAIC II, respectively.  The upper panel shows
the number of monitored stars per magnitude bin for all observed fields
parameterized by luminosity function. The cumulative number of
observed stars is then obtained by integrating over magnitude (see
lower panel). The trial LF's $\Phi_>$, $\Phi_A$, $\Phi_B$, $\Phi_C$,
and $\Phi_<$, are described in the text.
\label{fig:Nobs}}
\end{figure}

First, we estimate the number of observed stars $N_{obs}$ for each
field and amplifier using Equation~\ref{equ:Nobs} as described in
\S~\ref{sec:calcNobs}. Figure~\ref{fig:Nobs.sme9} illustrates this for
the example field sme9. The upper panel shows the range of luminosity
functions we considered.  The middle panel shows the detection efficiency\footnote{The
efficiency levels out at about 50\% since we integrate in
Equation~\ref{equ:eps} from $0$ to only $0.5$, and not to $1$, due to
the fact that we do not consider events with amplification smaller
than $2$}, and the lower panel shows the number of useful observed
stars per magnitude, which is the product of the two upper panels. 
Two competing effects, increasing star counts but decreasing efficiency for
fainter magnitudes, cause a peak at $M \sim -7.6$. Stars in the magnitude
range $-9<M<-6$ will contribute the most to the observed microlensing
event rate. Using a zeropoint magnitude of $31.3$ for this image, this
corresponds to a $VR$ filter (unlensed) magnitude range of $22.3 < M_{VR} <
25.3$.  The upper panel of Figure~\ref{fig:Nobs} shows the
sum of the number of observed stars per magnitude bin for all fields and
amplifiers, after correcting for exposure times. 

We obtain the total number of observed stars by
integrating over magnitude (see lower panel). Not surprisingly, the LF
chosen does significantly impact the estimated number of observed
stars, e.g., choosing $\Phi_B$ (short-dashed red line) produces
$\sim 2$ times more observed stars than $\Phi_C$ (dot-short-dashed
blue line). The temporal completeness is taken into account by using
$\tempcompl \sim 0.8$ (see \S~\ref{sec:tempcompl}).

We can now apply Equation~\ref{equ:Nml} to calculate $N_{ml}$ for any
combination of field, amplifier, and optical depth $\tau$. We use 
$\tau_{screen}=1.2 \times 10^{-7}$ for the optical depth for screen--lensing,  as determined by
the MACHO Project
\citep[][]{Alcock00a}. For self-lensing, we determine the 
optical depth for the Zhao \& Evans
self-lensing model sets {\it A} and {\it B} (see
\S~\ref{sec:selflensingmodels}) using Equation~\ref{equ:tauself}.  The 
field-dependent $\tau_{self}$ for each of these models is calculated using
each possible combination of $f_b=[0.3,0.4,0.5]$, and
$Z_0=[-1,-0.5,0,0.5,1]$, where $f_b$ is the mass fraction of the bar,
and $Z_0$ is the level of displacement between the disk and the bar in
kpc. Based on observations, this covers the likely parameter space of
$f_b$ and $Z_0$. The fields are divided into sets 1-5, based on their
respective star density determined with the \cite{Zhao00a} LMC bar
model (see Figure~\ref{fig:smfields}). Set 1 is the most crowded
(yellow) and set 5 is the least crowded (green). For each set of
fields, we add up the predicted number of microlensing events for the
different models and LF's (see Table~\ref{tab:eventrates}). The upper
panel of Figure~\ref{fig:ml2set} shows for each set of fields the
number of microlensing events for Zhao LMC self-lensing model set {\it
A} (black circles), model set {\it B} (blue squares), and
screen-lensing (red triangle). For clarity the open symbols are
plotted with a slight offset for a given set. Note that the spread in
the anticipated number of detected 
microlensing events for screen-lensing is solely due to the different
LF's ($\Phi_A$, $\Phi_B$, and $\Phi_C$) used (see
Figure~\ref{fig:LFs2sme9}). For self-lensing, an additional source of
spread is caused by using different values of $f_b$ and $Z_0$. For
sets 1-3 the intrinsic difference in event rate for self- and
screen-lensing is of the same order as the systematic errors.  The
event rate for self-lensing at the center of the bar is particularly
large if the displacement between the disk and the bar is large
($\vert Z_0 \vert=1$). For the two outer sets, the rate of
self-lensing strongly decreases and is well below the event-rate for
screen-lensing.

\section{Discussion}
\label{sec:discussion}

\begin{figure}[t]
\epsscale{0.9}
\plotone{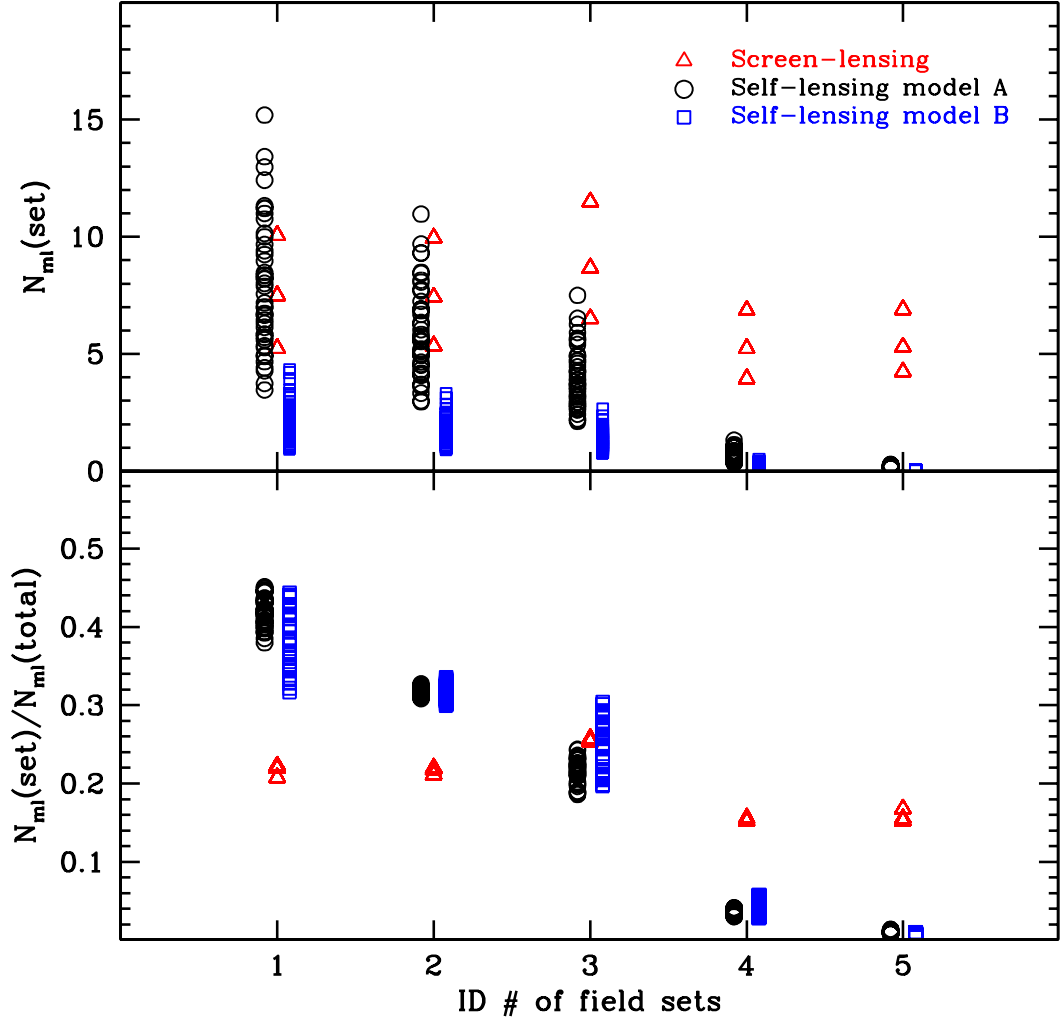}
\caption[Microlensing event rate and differential rate per field set]{
Estimated microlensing event counts and differential event rates for
the different field sets, as shown in Figure~\ref{fig:smfields}.  The
red triangles indicate screen-lensing for the different LF's.  The
Zhao \& Evans LMC self-lensing model sets {\it A} and {\it B} are
indicated with black circles and blue squares, respectively. The
disk-bar displacement of the LMC models varies between -1 and 1 kpc,
and the bar mass fraction varies between 0.3 and 0.5. For clarity the
open symbols are plotted with a slight offset for a given set.  The
upper panel shows the number of events over a 5 year
survey. The lower panel shows the differential event rate, normalized
to the total number of events detected over the course of the survey.
Note that the dependence on the LF's is greatly reduced for the
differential rates, as shown in the lower panel. In particular, by
comparing the innermost to the outermost fields we expect to be able
to distinguish between the screen-lensing and LMC self-lensing
scenarios.
\label{fig:ml2set}}
\end{figure}

Generally, the intrinsic difficulty with drawing conclusions from the
event rate is that there is a large spread in the predicted rate for a
given field due to systematic biases, e.g. the LMC or Galaxy model
used and the shape of the LF especially at the faint end. A way out of
this dilemma is to go from an absolute measurement to a relative
measurement: instead of considering absolute event rates we
investigate the differential event rate for a given field, defined as
the ratio of event rate of the field to the total event rate of all
fields (see lower panel of Figure~\ref{fig:ml2set}). Using this
normalized quantity to characterize the lensing rate across the LMC
suppresses the dependence on luminosity function. This is clearly
indicated in comparing the upper to the lower panels in
Figure~\ref{fig:ml2set}.  To zeroth order the systematic error arising
from LF uncertainty cancels out, and the measurement is much more
robust. Table~\ref{tab:eventrateratios} shows the differential event
rates for self- and screen-lensing.

Despite the anticipated increase in the number of microlensing events
for the SuperMACHO survey, we are nevertheless faced with using small
number statistics to try to distinguish between models for the lensing
population.  The basic approach we use is to consider what underlying
rate could be statistically consistent with a given observed number of
events.

We represent these results in confidence level plots, for which we use
the statistics given in \cite{Gehrels86}. Because the differences
between self- and screen-lensing are most pronounced in the outer
field sets, we will investigate field set 5 and then the combined
field sets 4 and 5.

\begin{figure}[t]
\epsscale{0.9}
\plotone{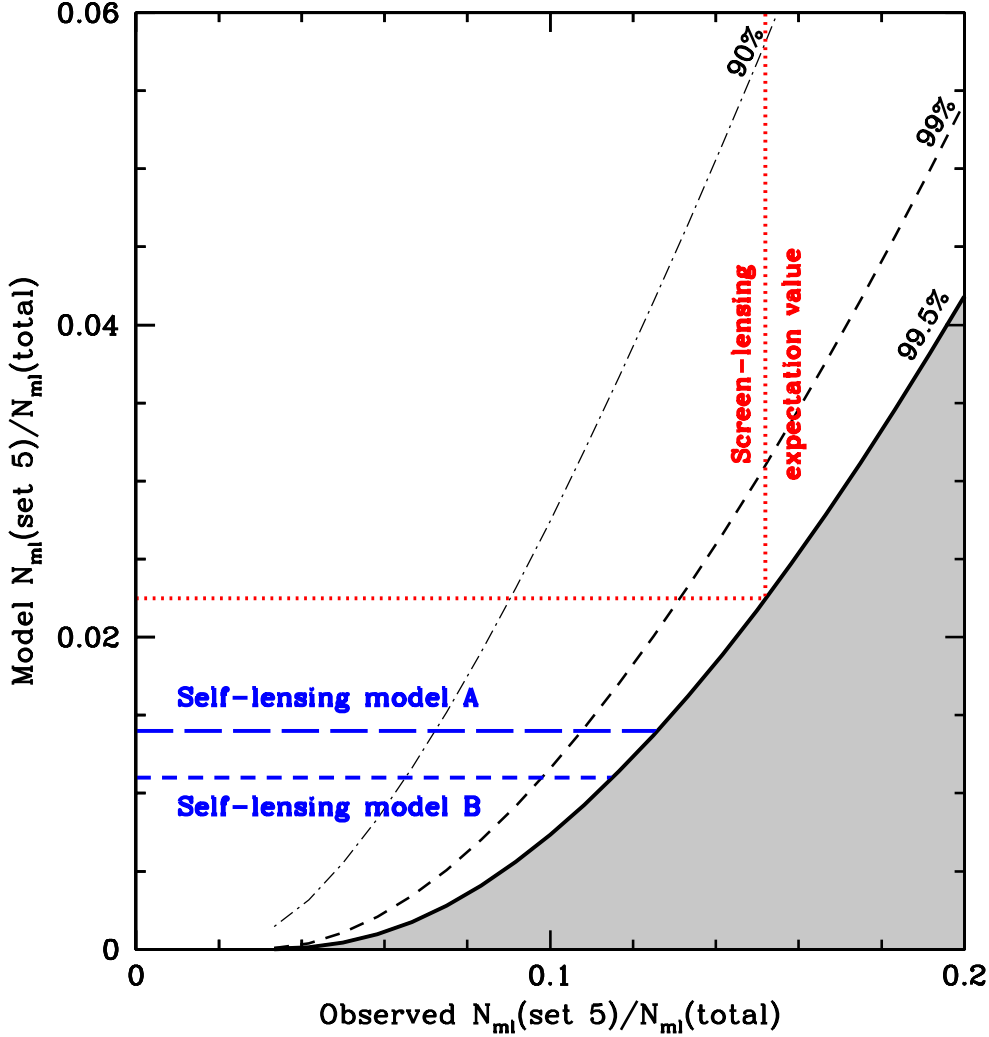}
\caption[Lower limit confidence plots for differential microlensing
event rates]{ 
Model exclusion plots for any observed differential microlensing event
rate.  The x--axis shows potential observed differential event rates for
SuperMACHO field set 5.  The y--axis corresponds to the allowed underlying
event rate that SuperMACHO's observations will use to constrain models.  The
upper limits to rates allowed by the observations at the 90\%,
99\%, and 99.5\% confidence levels are indicated with dot--dashed,
dashed, and solid black lines, respectively, based on Poisson
statistics assuming a total of 30 detected events.
For a given observed differential rate (as plotted on the x--axis),
the vertical projection of this measurement in the Figure will
intersect the various confidence contours.  The horizontal projection
of this intersection to the y--axis yields the {\it maximum}
differential ratio allowed by the observations.  In particular, if the
prior expectation value for screen lensing of 0.15 is actually
observed by SuperMACHO, we can exclude with 99.5\% confidence any model
that predicts a ratio of 0.023 or less.  This includes the self-lensing
models A (0.014) and B (0.011) as described in the paper.
\label{fig:lowerlimitratio}}
\end{figure}

Figure~\ref{fig:lowerlimitratio} is an illustration of the outcome of
this process. Assuming a total of 30 events are detected in the survey
(our estimate of the lower bound we're likely to see for screen
lensing), the x-axis corresponds to possible {\it observed}
differential event rates in field set 5.  The y-axis indicates the
lowest allowable {\it actual} underlying differential rate, given
Poisson statistics. The contours show the 90\% (dot-dashed black
line), 99\% (dashed black line), and 99.5\% (solid black line)
confidence limits on the minimum actual underlying rate, given some
measurement on the x-axis. The gray area shows the region excluded at
99.5\% confidence.

How can we use this plot? As a {\it Gedankenexperiment}, let us assume
that screen-lensing is indeed the underlying mechanism. In that case,
the expectation value for the field set 5 differential event rate is
0.15 for the least favorable LF (see Column~3 in
Table~\ref{tab:eventrateratios}).  Any particular experiment will
measure a rate different than exactly 0.15, but it will most likely
realize a value somewhere between 0.1 and 0.2.  As a guide, this
screen-lensing expectation value is indicated with a red dotted line
in Figure~\ref{fig:lowerlimitratio}.  The question is, can these
measurements exclude self-lensing?  The upper limit of the
differential rate of all Zhao self-lensing models in set {\it A} is
0.014 (see Column~5 in Table~\ref{tab:eventrateratios} and the blue,
long-dashed line in the confidence plot).  We can exclude these models
with 99\% and 99.5\% confidence {\it if} we observe a differential
rate larger than 0.11 and 0.13, respectively.  Similarly for model set
{\it B}, the maximum differential rate is 0.011 (blue short-dashed
line), thus it can be excluded at 99.5\% confidence if the observed
differential event rate is bigger than 0.12.  If SuperMACHO measures a
field set 5 differential rate larger than 0.13, than we can exclude
self-lensing as the dominant mechanism at the 99.5\% confidence level.

Let us consider the other case: If self-lensing is the dominant
lensing process, then we expect to find many events in the central
fields, and at most a couple in the outer fields. Therefore in this
case it makes sense to calculate the upper bound of the allowed
underlying differential rate given the observed number of events in
the outer fields and in total.

\begin{figure}[t]
\epsscale{0.9}
\plotone{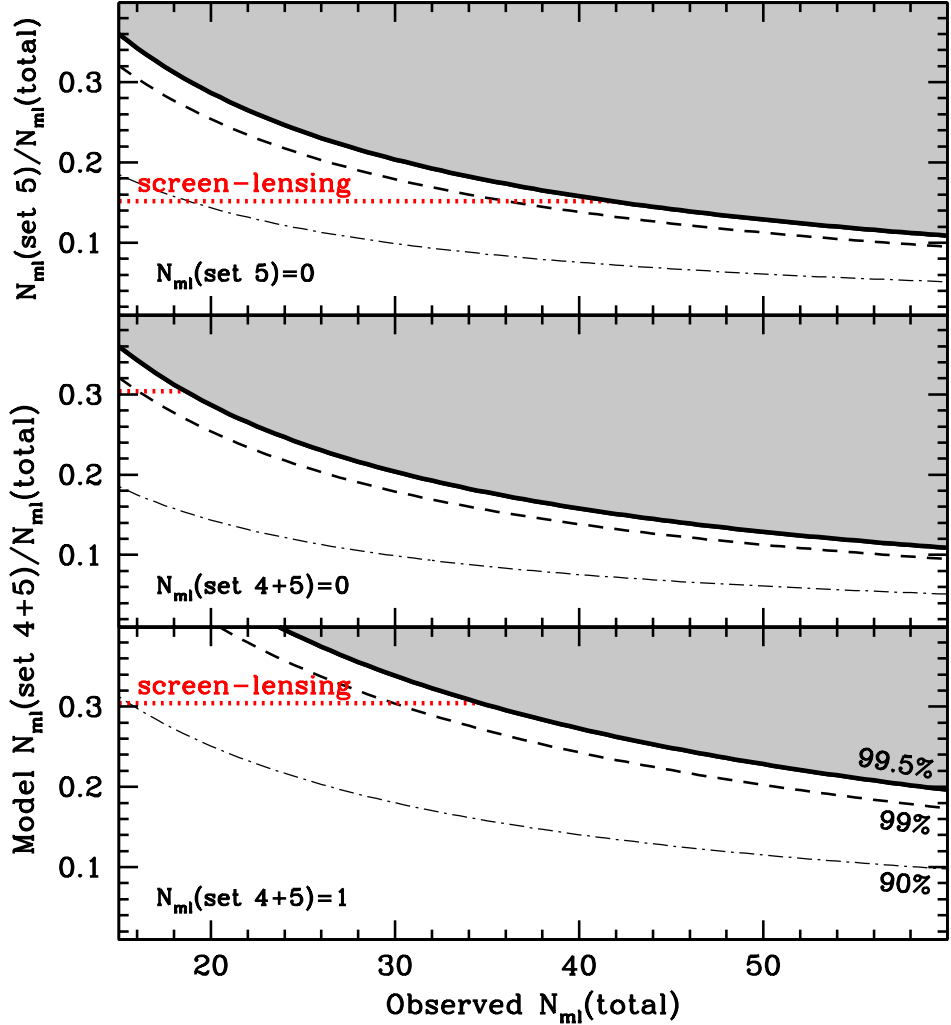}
\caption[Upper limit confidence plots for differential microlensing event rates]{ 
SuperMACHO's model discrimination ability as a function of the total
number of detected microlensing events.  The three panels correspond
to different possible experimental outcomes (0 or 1 events in the
outermost fields).  The x--axis shows the total number of detected
events.  The y--axis shows the true differential event rate in fields
4 + 5, and in field 5 only.  The lower boundary of the areas allowed
by the observations at the 90\%, 99\%, and 99.5\% confidence levels
are indicated with the dotted, dashed, and solid black lines,
respectively.  For a given observed total number of microlensing
events (x--axis) and the conditions listed inside each Figure window,
the confidence contours show the maximum underlying differential rate
allowed by the observations.  For example, in the case where no event
is observed in field sets 4 and 5 (middle panel), if 18 total events
are observed, any model predicting an underlying differential event
rate (set 4+5) greater than 0.3 (for example screen-lensing, see red,
dashed line) is excluded at 99.5\% confidence.
\label{fig:upperlimitratio}}
\end{figure}

For example, the most optimistic self-lensing model {\it A} predicts
0.28 events in field set 5 in 5 years. Most likely we will find either
0 or 1 event in this field set (assuming self-lensing only). 
If we find more than 18 events in total none of which are in field set
5, then we can exclude screen-lensing (assuming a differential rate of
0.17, see red, dashed line in the upper panel of
Figure~\ref{fig:upperlimitratio}) at the 90\% confidence level.  Note
that our self-lensing models predict a wide range of event rates from
5 to 35 events total during the 5 year survey (see
Table~\ref{tab:eventrates}), thus we can only expect to find that many
self-lensing events if the LMC is similar to the models with strong
central self-lensing. We would need 42 events to exclude
screen-lensing at the 99.5\% level and none in field set 5.

If of 18 or more total events we observe none in field sets 4 {\it
and} 5 (where screen lensing has an expectation rate of 0.30, see red,
dashed line in middle panel of Figure~\ref{fig:upperlimitratio}), we
can exclude screen-lensing with a confidence level greater than
99.5\%.  Finding no event is not unlikely since for the combined field
set 4 and 5, the event rates predicted from our self-lensing models
are between 0.25 and 1.6 events during the five years (see
Table~\ref{tab:eventrates}). Finding even one event in field sets 4
and 5 will weaken these conclusions, which can be seen by comparing
the exclusion regions of the middle panel to those of the lower panel.

In reality, the microlensing is probably not caused by a single lens
population, but rather by a mixture of several populations. We can
expect that it will be more difficult to differentiate between the
populations.  Still, very recent work re-investigating MACHO and EROS2
events finds that even though some of the events are due to
self-lensing, the total event rate and spatial distribution cannot be
explained by self-lensing alone \citep{Jetzer02}. Also, a large
spectroscopic survey targeting kinematic outliers in the LMC did not
find evidence for a significant additional, kinematically distinct,
population in the LMC \citep{Zhao02}. The lack of such a population
constrains the optical depth of self-lensing to values too small to
explain the observed event rate. Even if screen-lensing is the cause
for only a fraction of the observed event rate, we will be able to
detect enough events in the outer field sets to exclude self-lensing
as the sole lensing mechanism toward the LMC. If self-lensing is
excluded as the sole lensing mechanism, the SuperMACHO event rate will
provide a lower limit on the number of MACHO's in the halo of the MW
and thus provide a lower limit of their contribution to the MW's dark
matter.

\section{Summary and Conclusions}

The reported microlensing event rate toward the LMC exceeds that
expected from known visible components of our Galaxy, and the source
of this observed excess rate is still the subject of discussion.
Determining the nature of the lens population will have a great impact
on our understanding of Galactic and LMC structure and possibly on the
nature of dark matter. Possible explanations for the observed lensing
can be broadly categorized into screen-lensing and self-lensing
scenarios.  Using the first year data of the SuperMACHO Project, we
performed completeness analysis by adding artificial stars to the
images and estimated the spatial and temporal efficiency of detecting
microlensing events in our data sets. We predicted a lower limit of
observable microlensing events for both categories using the
efficiency in combination with the stellar luminosity functions.  We
find that the SuperMACHO Project should be able to distinguish between
the two categories using the spatial differences in optical
depth. Utilizing the differential event rate instead of the event rate
itself decreases the impact of systematic errors rendering the results
and conclusions more robust.

\section{Acknowledgments}

The SuperMACHO survey is being undertaken under the auspices of the 
NOAO Survey Program. We are very grateful for the support 
provided to the Survey program from the NOAO and the National Science Foundation
We are particularly indebted to the scientists and staff at the Cerro
Tololo Interamerican Observatory for their assistance in helping us carry
out the SuperMACHO survey. We also appreciate the invaluable help of Mr. Chance
Reschke in building and maintaining the computing cluster we use for 
image analysis. This project works closely with members of the ESSENCE
supernova survey, and we are grateful for their input and 
assistance.

The support of the McDonnell Foundation, through a Centennial Fellowship 
awarded to C. Stubbs, has been essential to the SuperMACHO
survey. We are most grateful for the Foundation's support for this project.
Stubbs is also grateful for support from Harvard University.

KHC's, SN's and GP's work was performed under the auspices of the U.S.
Department of Energy, National Nuclear Security Administration by the
University of California, Lawrence Livermore National Laboratory under
contract No. W-7405-Eng-48.

DLW acknowledges financial support in the form of a Discovery 
Grant from the Natural Sciences and Engineering Research Council 
of Canada (NSERC).


\bibliographystyle{apj}
\bibliography{ms}

\clearpage


\clearpage

\clearpage

\clearpage

\clearpage

\clearpage

\clearpage

\clearpage

\clearpage

\clearpage

\clearpage


\begin{deluxetable}{ccccccc}
\tablecaption{Luminosity Function Parameterization 
\label{tab:LFoverview}}
\tablehead{
\colhead{LF} & \colhead{$\Phi_1$} & \colhead{$\beta_1$} &
\colhead{$\beta_2$} & \colhead{$\beta_3$} & \colhead{$M_1$} &
\colhead{$M_2$}\\
(1) & (2) & (3) & (4)  & (5) & (6) & (7) 
}
\startdata
$\Phi_>$   & $\Phi_0$  & $\beta$ & - & - & - & - \\
$\Phi_{A}$ & $\Phi_{A0}$ & $0.34$ & $0.16$  & $0.027$  & $M_{RC} + 4.3$ & $M_{RC} + 7.5$ \\
$\Phi_{B}$ & $\Phi_{B0}$ & $0.38$ & $0.031$ & -       & $M_{RC} + 5.0$  & - \\
$\Phi_{C}$ & $\Phi_{0}$ & $\beta$ & $0.16$  & $0.027$  & $M_{RC} + 3.0$  & $M_{RC} + 7.5$ \\
$\Phi_<$ & $\Phi_{0}$ & $\beta$ & $0.027$  & -       & $M_{RC} + 3.0$  & - \\
\enddata
\tablecomments{
Overview of the different luminosity function (LF) parameterizations
used for computing the number of monitored sources, in the terminology
used in the text. Column (1) shows the name of the LF model. Column
(2) indicates the stellar density parameter used. Columns (3)-(5)
indicate the relevant power law slopes, if applicable. Columns (6) and
(7) show the transition magnitude between power-laws with respect to
the fitted red clump magnitude $M_{RC}$.  }
\end{deluxetable}

\begin{deluxetable}{ccccc}
\tablecaption{Predicted Microlensing Event Rates for the SuperMACHO Survey
\label{tab:eventrates}}
\tablehead{
\colhead{Lens Pop.} & \colhead{Field Set} & \colhead{Min} & \colhead{Mean}  & \colhead{Max}\\
(1) &  (2) & (3) & (4) & (5) \\
}
\startdata
screen-lensing & 5 & 4.2 & 5.5 & 6.9 \\
& 4+5 &  8.2 & 10.8 & 13.8\\
& all & 25.2 & 34.9 & 45.2 \\
\hline
LMC self-lensing & 5 & 0.12 & 0.18 & 0.28\\
 model set {\it A} & 4+5 & 0.47 & 0.87 & 1.6 \\
& all & 9.1 & 18.7 & 35.3 \\
\hline
LMC self-lensing & 5 & 0.03 & 0.07 & 0.13 \\ 
 model set {\it B}& 4+5 & 0.25 & 0.53 & 1.13 \\
& all & 5.2 & 10.7 & 21.4 \\
\hline
\enddata
\tablecomments{
Predicted microlensing event totals for a 5 year SuperMACHO
survey. The lens population and field set are indicated in column (1)
and column (2), respectively. The minimum, mean, and maximum event
rates for the different models and LF's are shown in column (3), (4),
and (5), respectively. The Zhao \& Evans LMC self-lensing model set
{\it A} and {\it B} differ as described in
\S~\ref{sec:selflensingmodels}. Note that independent of the overall
rate normalization, the {\em ratios} of rates are a clear and robust
indicator of the nature of the lensing population.  }
\end{deluxetable}

\begin{deluxetable}{ccccc}
\tablecaption{Predicted Event Rate Ratios
\label{tab:eventrateratios}}
\tablehead{
\colhead{Lens Pop.} & \colhead{Field Set} & \colhead{Min} & \colhead{Mean}  & \colhead{Max}\\
(1) &  (2) & (3) & (4) & (5) \\
}
\startdata
screen-lensing & 5 & {\bf 0.15} & 0.16 & 0.17\\
& 4+5 & {\bf 0.30} & 0.31 & 0.32 \\
\hline
LMC self-lensing & 5 & 0.008 & 0.010 & {\bf 0.014}\\
model set {\it A}  & 4+5 & 0.038 & 0.047 & {\bf 0.055} \\
\hline
LMC self-lensing & 5 & 0.004 & 0.007 & {\bf 0.011}\\
model set {\it B}  & 4+5 & 0.032 & 0.051 & {\bf 0.070} \\
\enddata
\tablecomments{
Predicted differential microlensing event rates for screen and
self-lensing. The lens population and field set is indicated in column
(1) and column (2), respectively. The minimum, mean, and maximum
differential rates for the different models and LF's are shown in
column (3), (4), and (5), respectively.  In order to be conservative,
we use the minimum values for screen-lensing and the maximum values
for self-lensing, indicated in bold.  The Zhao \& Evans LMC
self-lensing model set {\it A} and {\it B} differ as described in
\S~\ref{sec:selflensingmodels}.  Note that independent of the
overall rate normalization, the {\em differential} rates are a clear
and robust indicator of the nature of the lensing population.
}
\end{deluxetable}

\end{document}